\documentclass[epj]{svjour}
\usepackage{graphicx,color,amsmath, amssymb,cite,multirow}
\usepackage[english]{babel}
\usepackage{t1enc}
\usepackage[latin2]{inputenc}
\begin{document}

\title{Observables from a solution of 1+3 dimensional relativistic hydrodynamics}

\author{Máté Csanád\inst{1} \and Márton Vargyas\inst{1}}
\institute{Eötvös University, Department of Atomic Physics, Pázmány Péter s. 1/A, H-1117 Budapest, Hungary}

\date{\today}
\abstract{In this paper we analyze a 1+3 dimensional solution of relativistic hydrodynamics.
We calculate momentum distribution and other observables from the solution and
compare them to measurements from the Relativistic Heavy Ion Collider (RHIC). We find that the
solution we analyze is compatible with the data. In the last several years many numerical
models were tested, but it is the first time that an exact, parametric, 1+3 dimensional relativistic
solution is compared to data.}

\maketitle

\section{Introduction}
In the last several years it has been revealed that the matter produced in the collisions of the Relativistic Heavy Ion Collider (RHIC)
is a nearly perfect fluid~\cite{Lacey:2006bc}, i.e.\ it can be described with perfect fluid hydrodynamics. There was a long search for exact hydrodynamic solutions
(solutions of the partial differential equations of hydrodynamics) and several solutions proved to be applicable. There are 1+3 dimensional solutions,
 as well as relativistic solutions - but no 1+3 dimensional \emph{and} relativistic exact solution has been tested yet. In this paper
we extract observables from the relativistic, ellipsoidally symmetric solution of ref.~\cite{Csorgo:2003ry}. We calculate momentum distribution,
elliptic flow and correlation radii and compare them to RHIC data.

\section{Perfect fluid hydrodynamics}
Perfect fluid hydrodynamics is based on local conservation of
entropy or number density ($n$), energy-momentum density ($T^{\mu\nu}$), expressed by so-called conservation equations:
\begin{align}
\label{e:rhd}
		\partial{_\mu}(nu^\mu)&=0,\\
		\partial{_\mu}T^{\mu\nu}&=0,
\end{align}
where $u^\mu$ is the flow field in the fluid. The fluid is perfect if the energy-momentum tensor is diagonal in the local rest frame, i.e.\ 
viscosity and heat conduction are negligible. This can be assured if $T^{\mu\nu}$ is chosen as
\begin{align}
T^{\mu\nu}=(\epsilon+p)u^\mu u^\nu-pg^{\mu\nu},
\end{align}
where $\epsilon$ is energy density, $p$ is pressure and $g^{\mu\nu}$ is the metric tensor, diag(1,-1,-1,-1).
The conservation equations are closed by the equation of state, which gives the relationship between $\epsilon$ and $p$. Typically $\epsilon = \kappa p$ is chosen,
where the proportionality ``constant'' $\kappa$ may depend on temperature $T$, which is connected to the density $n$ and pressure $p$ via $p=nT$. In some solutions
(as also on the analyzed one) a bag constant $B$ can be introduced (this is not favored by the data, because no first order phase transition is seen in high energy heavy
ion collisions, these are in the cross-over regime of the QCD phase-diagram~\cite{Fodor:2009ax}). The exact, analytic result for hydrodynamic
solutions is, that the hadronic observables do not depend on the initial state or the dynamical equations separately, just through the final state~\cite{Csanad:2009sk}.
Thus if we fix the final state from the data, the equation of state can be anything that is compatible with the particular solution.
This is the framework of several hydro solutions as detailed in the next paragraph.

Even though many solve the above equations numerically, there are only a few exact solutions for these equations. One (and 
historically the first) is the implicit solution discovered more than 50 years
ago by Landau and Khalatnikov ~\cite{Landau:1953gs,Khalatnikov:1954aa,Belenkij:1956cd}. This
is a 1+1 dimensional solution, and has realistic properties: it
describes a 1+1 dimensional expansion, does not lack acceleration
and predicts an approximately Gaussian rapidity distribution.
Another renowned solution of relativistic hydrodynamics was found by
Hwa and Bjorken solution~\cite{Hwa:1974gn,Chiu:1975hw,Bjorken:1982qr}:
it is simple, 1+1 dimensional, explicit and exact, but
accelerationless. It is boost-invariant in its
original form. Boost invariance is however incompatible with data
from RHIC, so the solution fails to describe the data (it still can be used
to estimate the energy density in high energy heavy ion collisions; note that
this solution rather underestimating the initial energy density, see ref~\cite{Csorgo:2006ax}).
Important are solutions~\cite{Csorgo:2006ax,Bialas:2007iu} which are
explicit and describe a relativistic acceleration, i.e. combine the
properties of the Landau-Khalatnikow and the Hwa-Bjorken solutions. With these
one can have an advanced estimate on the energy density, but investigation of transverse
dynamics is not possible by these solutions.

There were no 1+3 dimensional relativistic exact parametric solutions investigated yet: the only exact solution is
the one in ref.~\cite{Csorgo:2003ry}. Observables from this solution were not computed and compared
to data yet. Present paper hence calculates observables from a realistic 1+3 dimensional solution and compares them to data for the first time.
There were several numerical solutions of relativistic hydrodynamics compared to data, but our method
is different from numerical calculations: here one can determine the best values of the parameters of
the solution by fitting the analytic results to data.

\section{The analyzed solution}
The analyzed solution~\cite{Csorgo:2003ry} assumes self-similarity and ellipsoidal symmetry. The ellipsoidal symmetry means
that at a given proper time the thermodynamical quantities are constant on the surface of expanding ellipsoids.
The ellipsoids are given by constant values of the scale variable $s$, defined as
\begin{align}
s=\frac{r_x^2}{X(t)^2}+\frac{r_y^2}{Y(t)^2}+\frac{r_z^2}{Z(t)^2},
\end{align}
where $X(t)$, $Y(t)$, and $Z(t)$ are time dependent scale parameters (axes of the $s=1$ ellipsoid),
only depending on the time $t$. Spatial coordinates are $r_x$, $r_y$, and $r_z$.
The velocity-field is described by a Hubble-type expansion:
\begin{align}
u^\mu (x) = \gamma \left(1, \frac{\dot X(t)}{X(t)}r_x, \frac{\dot Y(t)}{Y(t)}r_y, \frac{\dot Z(t)}{Z(t)}r_z\right),
\end{align}
where $x$ means the four-vector $(t,r_x,r_y,r_z)$, and $\dot X(t) = dX(t)/dt$, similarly for $Y$ and $Z$.
The $\dot X(t)=\dot X_0$, $\dot Y(t)=\dot Y_0$, $\dot Z(t)=\dot Z_0$ (i.e. all are constant) criteria must
be fulfilled, ie.\ the solution is accelerationless. This is one of the drawbacks of this solution.

The temperature $T(x)$ and number density $n(x)$ are:
\begin{align}
n(x)&=n_0\left(\frac{\tau_0}{\tau}\right)^3 \nu(s), \\
T(x)&=T_0\left(\frac{\tau_0}{\tau}\right)^{3/\kappa} \frac{1}{\nu(s)},\label{e:temp}\\
p(x)&=p_0\left(\frac{\tau_0}{\tau}\right)^{3(\kappa+1)/\kappa},
\end{align}
where $\tau$ is the proper time, $s$ is the above scaling variable, $\nu(s)$ is an arbitrary function, while $n_0=n|_{s=0,\tau=\tau_0}$, $T_0=T|_{s=0,\tau=\tau_0}$
and $p_0=p|_{s=0,\tau=\tau_0}$ with $p_0 = n_0 T_0$ (hence $p$ does not depend on the spatial coordinates only $\tau$). Furthermore, $\tau_0$ is an arbitrary proper time, 
but practically we choose it to be the time of the freeze-out, thus $T_0$ is the central freeze-out temperature.
Note that in our solution, the parameter $\kappa$ is arbitrary, i.e. any value of $\kappa$ yields a solution.
The function $\nu(s)$ is chosen as:
\begin{align}
\nu(s)=e^{-bs/2},
\end{align}
where $b = \left.\frac{\Delta T}{T}\right|_r$ is the temperature gradient. If the fireball is the hottest in the center, then $b<0$.

\section{Freeze-out and source function}
The picture widely used in hydro models is that the pre freeze-out (FO) medium is described by hydrodynamics, and the post FO medium is that
of observed hadrons. Note that cases have been analyzed where the pre and post
FO physical parameters are different, see refs.~\cite{Csernai:1995zn,Magas:2007zz,Csorgo:1994dd} for details. In our framework we assume however
that the freeze-out can happen at any proper time, e.g in case of a self-quenching effect or if the phase space evolution is that of a collisionless gas.
Thus there is no jump in the equation of state post and pre FO, i.e. $\kappa$ goes to $\kappa_{\rm free}$ smoothly,
to the EoS of free hadrons. This is a widely used assumption, the framework of our solution is similar to that of refs.~\cite{Csorgo:2001xm,Csorgo:2006ax,Nagy:2009eq}.

In this case the hadronic observables can be extracted from the solution via the phase-space distribution at the FO. This will
correspond to the hadronic final state or source distribution $S(x,p)$.

As mentioned in the previous section, one does not need to fix a special equation of state, because the same
final state can be achieved with different equations of state or initial conditions~\cite{Csanad:2009sk}. Thus in this
paper $\kappa$ is arbitrary -- the hadronic observables do not restrict the value of $\kappa$. The bag constant $B$ does
not have to be specified either, but the instantaneous FO we assume corresponds to zero bag constant.

In our solution the source distribution takes the following form:
\begin{align}
S(x,p)d^4x=  \mathcal{N}B(x,p) H(\tau)d\tau d^3 \Sigma_\mu(x) p_\mu,
\end{align}
where $\mathcal{N}=g/(2\pi)^3$ (with $g$ being the degeneracy factor), $H(\tau)$ is the proper-time probability
distribution of the FO, $B(x,p)$ is the Boltzmann-distribution and $d^3 \Sigma_\mu(x)p^\mu$ is the Cooper-Frye factor~\cite{Cooper:1974mv}
describing the flux of the particles, and $d^3 \Sigma_\mu(x)$ is the vector-measure of the FO hyper-surface. We assume that the FO
happens at a constant proper time $\tau_0$, i.e. $H(\tau)d\tau=\delta_{\tau_0}(\tau)$ and $d^3 \Sigma_\mu(x)=u^\mu d^3x/u^0$, i.e.
the FO hyper surface is assumed to be normal to $u^\mu$.

We assume for the Jüttner-distribution:
\begin{align}
B(x,p) = \exp \left[\frac{\mu(x) - p_\mu u^\mu(x)}{T(x)}\right]=n(x)\exp \left[-\frac{p_\mu u^\mu(x)}{T(x)}\right],
\end{align}
where $\mu(x)/T(x)=\ln n(x)+\mu_0/T_0$ is the fugacity factor.
 Finally the source distribution is: 
\begin{align}
S(x,p)d^4x=\mathcal{N}n(x)\exp \left[-\frac{p_\mu u^\mu(x)}{T(x)}\right]  \frac{p_\mu u^\mu}{u^0} \delta_{\tau_0}(\tau)d\tau d^3x.
\end{align}

Note that the source distribution is normalized such as $\int S(x,p) d^4 x d^3{\bf p}/E = N$,
i.e. one gets the total number of particles $N$ (using $c$=1, $\hbar$=1 units)

\section{Observables from the solution}
Now let us calculate the observables that are usually measured in high energy heavy ion collisions.
The invariant momentum distribution $N_1(p)$ can be calculated as:
\begin{align}
E\frac{d^3N}{d^3p} = N_1(p) = \int S(x,p) d^4 x.
\end{align}

For the integration, we did a second order Gaussian approximation.
The result for the invariant momentum distribution of a particle with mass $m$ is:
\begin{align}
 \label{e:n1p}
  N_1(p) =& \int S(x,p) d^4x =
   2 \pi \overline{N}\overline{E}\overline{V}\\
   \times &\exp \left[-\frac{E^2+m^2}{2ET_0}-\frac{p_x^2}{2ET_x}-\frac{p_y^2}{2ET_y}-\frac{p_z^2}{2ET_z}  \right],\nonumber
\end{align}
with the following auxiliary quantities:
\begin{align}
\overline{N}&=\mathcal{N} n_0 \left(\frac{2T_0\tau_0^2\pi}{E} \right)^{3/2},\\
\overline{E}&=\left(E-\frac{p_x^2 (1-\frac{T_0}{T_x})}{E}-\frac{p_y^2 (1-\frac{T_0}{T_y})}{E}-\frac{p_z^2 (1-\frac{T_0}{T_z})}{E} \right),\\
\overline{V}&=\sqrt{\left(1-\frac{T_0}{T_x}\right)\left(1-\frac{T_0}{T_y}\right)\left(1-\frac{T_0}{T_z}\right)}.
\end{align}
Furthermore, $T_x$ ,$T_y$, $T_z$ are the effective temperatures, i.e. inverse logarithmic slopes of the distribution:
\begin{align}
T_x&=T_0+\frac{E T_0\dot X_0^2}{b(T_0-E)}, \\
T_y&=T_0+\frac{E T_0\dot Y_0^2}{b(T_0-E)}, \\
T_z&=T_0+\frac{E T_0\dot Z_0^2}{b(T_0-E)},
\end{align}
where $\dot X_0$, $\dot Y_0$ and $\dot Z_0$ are the (constant) expansion rates of the fireball, $T_0$ its central temperature
at FO and $b$ the temperature gradient. There is an important criterion for the validity of this calculation: $T_{x,y,z}>T_0$ has
to be true. This is the case if $b<0$ (i.e. the fireball is cooler on the outside than in the inside) and $E>T_0$. In case of
roughly 200 MeV central freeze-out temperature, this yields $p>140 MeV/c$. We will see later, that this is true for all the data points we use.

\subsection{The transverse momentum distribution and the elliptic flow}\label{ss:n1v2}
The invariant momentum distribution depends on all three momentum coordinates. Here we take the longitudinal momentum to be zero,
as the data we compare it to are measured at mid-rapidity. We use transverse polar coordinates $\phi$ and $p_t$ instead of $p_x=p_t \cos(\phi)$ and $p_y=p_t \sin(\phi)$.
This way $N_1(p)$ can be rewritten as
\begin{align}
N_1(p)=N_1(p_t)\left[1+2\sum_{\substack{n=1}}^{\infty} v_n \cos(n\phi)\right],
\end{align}
where $v_n$ are the flow coefficients, in particular $v_2$ is the elliptic flow.
This way we may calculate the transverse momentum distribution $N_1(p_t)$ from $N_1(p)$ of eq.~(\ref{e:n1p}):
\begin{align}
N_1(p_t)=\frac{1}{2\pi}\int_{0}^{2\pi} N_1(p) d\phi
\end{align}
and the elliptic flow:
\begin{align}
v_2(p_t)=\frac{\int_{0}^{2\pi} d\phi N_1(p) \cos(2\phi)}{\int_{0}^{2\pi}d\phi N_1(p)}.
\end{align}

Result for the transverse momentum spectrum is:
(we substitute $E=m_t=\sqrt{m^2+p_t^2}$ as being at mid-rapidity):
\begin{align}
\label{e:n1pt}
  N_1(p_t)= &2\pi\overline{N}\;\overline{V} \left(m_t-\frac{p_t^2(T_{\rm eff}-T_0)}{m_tT_{\rm eff}}\right)\\
   \times& \exp\left[-\frac{m_t^2+m^2}{2m_tT_0}-\frac{p_t^2}{2m_tT_{\rm eff}}\right],\nonumber
\end{align}
where we introduced $1/T_{\rm eff}=0.5(1/T_x+1/T_y)$, the effective temperature.
The result for the elliptic flow is:
\begin{align}
\label{e:v2}
v_2(p_t)=\frac{I_1(w)}{I_0(w)}
\end{align}
where $I_0$, and $I_1$ are the modified Bessel functions while
\begin{align}
w=\frac{p_t^2}{4m_t}\left(\frac{1}{T_y}-\frac{1}{T_x}\right).
\end{align}
The formula for $v_2$ gives back previously found formulas of non-relativistic solutions~\cite{Csorgo:2001xm} and relativistic
solutions~\cite{Csanad:2003qa,Csanad:2005gv}. Also the formula for $N_1(p_t)$ is similar to results of the previously mentioned papers.

An important consequence of the above results is that neither $N_1$ nor $v_2$ depends on the EoS itself, only through the final state
parameters. If we determine for example $T_0$, the freeze-out central (at the center means here $r_x=r_y=r_z=0$) temperature,
$\kappa$ or the initial temperature $T_{\rm initial}$
still cannot be calculated. We only know that $T_{\rm initial} = T_0 (\tau_0/\tau_{\rm initial})^{3/\kappa}$, see eq.~(\ref{e:temp}).
Thus $\kappa$ or $T_{\rm initial}$ has to be determined from another measurement, e.g. the spectrum of thermal photons.

\subsection{The Bose-Einstein correlations}
The two-particle Bose-Einstein correlation function~\cite{Weiner:2000wj} is also calculable from the source distribution
\begin{align}
\label{e:c2sf}
C_2(q,K)=1+\left|\frac{\widetilde S(q,K)}{\widetilde S(q=0,K)}\right|^2,
\end{align}
with $q$ being the momentum difference and $K$ the average momentum of the pair (and $q\ll K$, hence $p_1\approx p_2 \approx K$), and $\widetilde S(q,K)$ is the Fourier transformed ($x\rightarrow q$) of $S(x,K)$.
The result of the solution is:
\begin{align}
\label{e:c2}
C_2(q,K)=1+\exp\left[-R_x^2q_x^2-R_y^2q_y^2-R_z^2q_z^2\right],
\end{align}
where $R_x, R_y, R_z$ are the correlation radii, called HBT radii:
\begin{align}
R_x^2&=\frac{T_0\tau_0^2(T_x-T_0)}{M_tT_x},\\
R_y^2&=\frac{T_0\tau_0^2(T_y-T_0)}{M_tT_y},\\
R_z^2&=\frac{T_0\tau_0^2(T_z-T_0)}{M_tT_z},
\end{align}
where $M_t$ is the transverse mass  belonging to the average momentum $K=0.5(p_1+p_2)$, which is (at mid-rapidity) $M_t=0.5\left(m_{t,1}+m_{t,2}\,\right)$. The
$m_{t,1}$, and $m_{t,2}$ quantities are the transverse masses, the $T_{x}$, $T_{y}$, and $T_{z}$ are the effective temperatures belonging to the average momentum (i.e.\ 
here $T_x=T_x|_{M_t}$). Note that the above formulas resemble the usual scaling of the HBT radii: $R^2 \propto 1/M_t$. This means that all radii scale with the average
transverse mass, independently of particle type, i.e.\ if plotted versus average $m_t$, kaon radii and pion radii fall onto the same scaling curve. Note that HBT radii
depend also only on final state parameters, not on the EoS itself, similarly to the elliptic flow or spectra (see the last paragraph of subsection~\ref{ss:n1v2} on this
matter).

To compare the HBT radii with the data the Bertsch-Pratt~\cite{Pratt:1986cc} frame is to be used. It has three axes: the \emph{out} is the direction of the average 
transverse momentum of the pair, the \emph{long} direction is equal to the direction \emph{z}, and the \emph{side} direction is orthogonal to both of them.
The result for $R_{\rm out}$, $R_{\rm side}$ and $R_{\rm long}$ is:
\begin{align}
R_{\rm out}^2=R_{\rm side}^2&=\frac{R_x^2+R_y^2}{2}, \label{e:ros}\\
R_{\rm long}^2&=R_z^2.
\end{align}
Clearly in this solution the out and side radii are equal. This can be attributed to the instantaneous freeze-out; a non-zero freeze-out duration would make
$R_{\rm out}^2$ bigger by a term of $\Delta \tau^2 p_t^2/E^2$. Supported by the data, we use the $\Delta \tau = 0$ approximation in our solution,
which corresponds to instantaneous freeze-out.

\section{Comparing the observables to RHIC data}
We compared the above results to PHENIX data of 200 GeV Au+Au collisions. We fitted our above formulas to spectra and HBT positive pion
data~\cite{Adler:2003cb,Adler:2004rq} (0-30\% centrality) and elliptic flow data~\cite{Adler:2003kt} for $\pi^\pm$, $K^\pm$, p and
$\overline{\rm p}$ particles (0-92\% centrality).

The used parameters for spectra and HBT are $T_0$ (central freeze-out temperature),
$\tau_0$ (freeze-out proper-time), $b$ (temperature gradient) and the expansion rates $\dot X_0$, $\dot Y_0$ and $\dot Z_0$. Instead of $\dot X_0$ and $\dot Y_0$ however
we use two more commonly used parameters, freeze-out expansion aniso\-tropy $\epsilon$ and average transverse expansion rate $u_t$:
\begin{align}
\epsilon&=\frac{\dot X^2 - \dot Y^2}{\dot X^2 + \dot Y^2},\\
u_t^2&=\frac{1}{2}\left(\frac{1}{\dot X^2}+\frac{1}{\dot Y^2}\right)
\end{align}.
It turns out however, that the expansion rates appear only in combination with
the temperature gradient $b$, hence we use $u_t^2/b$ and $\dot Z_0^2/b$ as fit parameters. In eq.~(\ref{e:v2}), the formula for the elliptic flow,
$\tau_0$ and $\dot Z^2_0$ do not appear, hence we do not use them when fitting $v_2$ data.

For analyzing the confidence levels we calculate $\chi^2$ and number of degrees of freedom (NDF) of the fits. However, in the calculation there are
numerous approximations. The main approximation is the Gaussian one in the integration. Here we neglect terms of higher order. Later, with the given
parameters we can estimate their contribution at a given point. The error coming from the integration can thus be estimated to be at least 3\%. In
table~\ref{t:pars}.\ we also give the $\chi^2$ using this additional error.

\begin{table}
	\centering
		\begin{tabular}{|c|c|c|}\hline
		  fit parameter   & $N_1$ and HBT & elliptic flow  \\ 
		                  &  0-30\% cent. &   0-92\% cent. \\\hline
		  $T_0$ [MeV]     & 199$\pm$3     & 204$\pm$7      \\
		  $\epsilon$      & 0.80$\pm$0.02 & 0.34$\pm$0.03  \\
		  $u_t^2/b$       & -0.84$\pm$0.08& -0.34$\pm$0.01 \\
		  $\tau_0$[fm$/c$]& 7.7$\pm$0.1   & -              \\
		  $\dot{Z}_0^2/b$ & -1.6$\pm$0.3  & -              \\\hline
		  NDF             & 41            & 34             \\
		  $\chi^2$        & 171           & 256            \\\hline
		  $\chi^2$ with 3\%&\multirow{2}{*}{24}&\multirow{2}{*}{66}  \\
		    theory error  &               &                \\ \hline
		\end{tabular}
	\caption{The parameters obtained from fits to Au+Au PHENIX data~\cite{Adler:2003cb,Adler:2003kt,Adler:2004rq}. The difference of the parameters can be
	explained by the centrality of the datasets (0-30\% for spectra and HBT versus 0-92\% for elliptic flow). The fact that $u_t^2/b$ is negative
	means that $b<0$, i.e.\ the fireball is the hottest in the center and colder outside (i.e.\ a Gaussian temperature profile).}
	\label{t:pars}
\end{table}

\section{Discussion of the results}
\begin{figure}
	\centering
		\includegraphics[height=0.48\textwidth,angle=270]{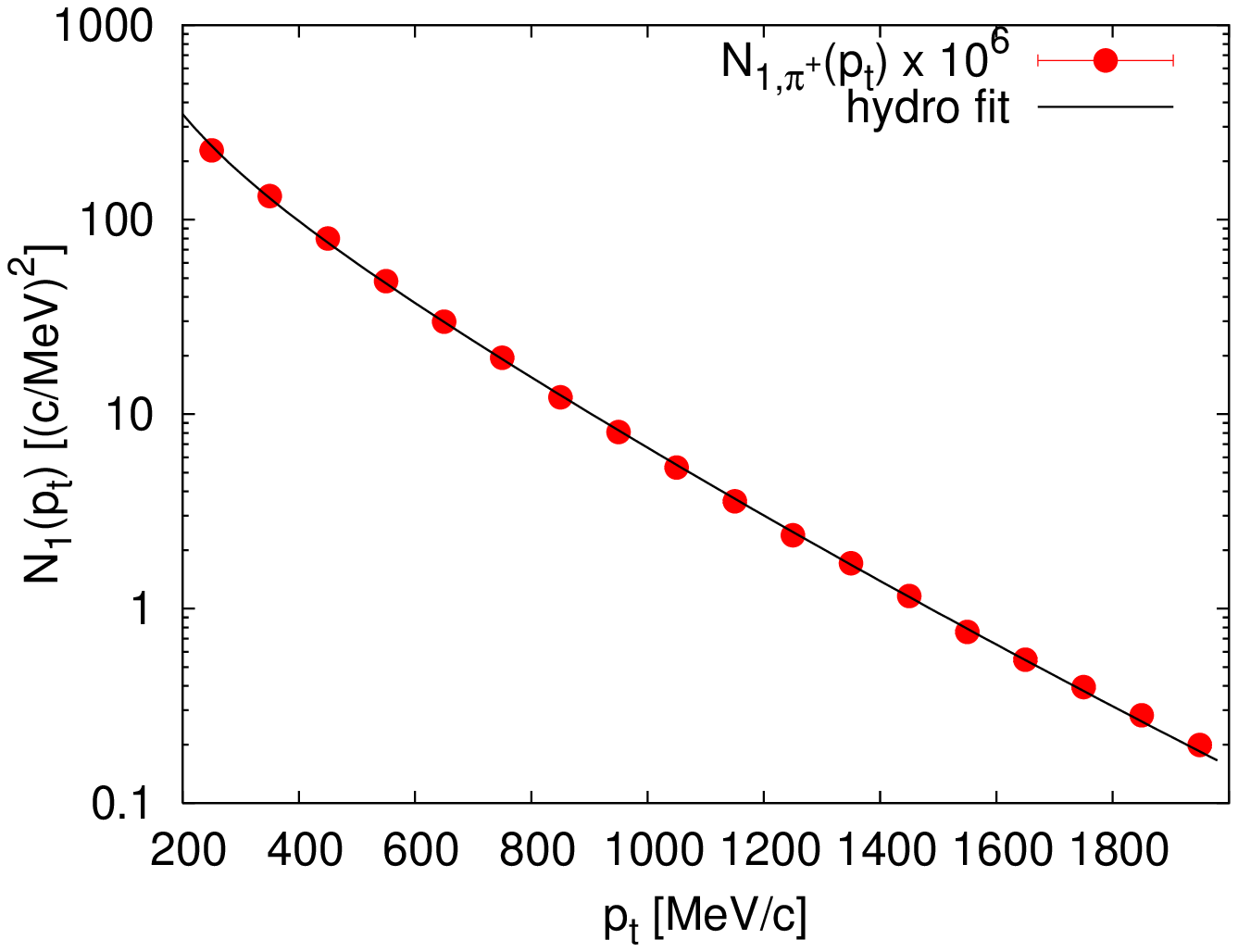}
		\includegraphics[height=0.48\textwidth,angle=270]{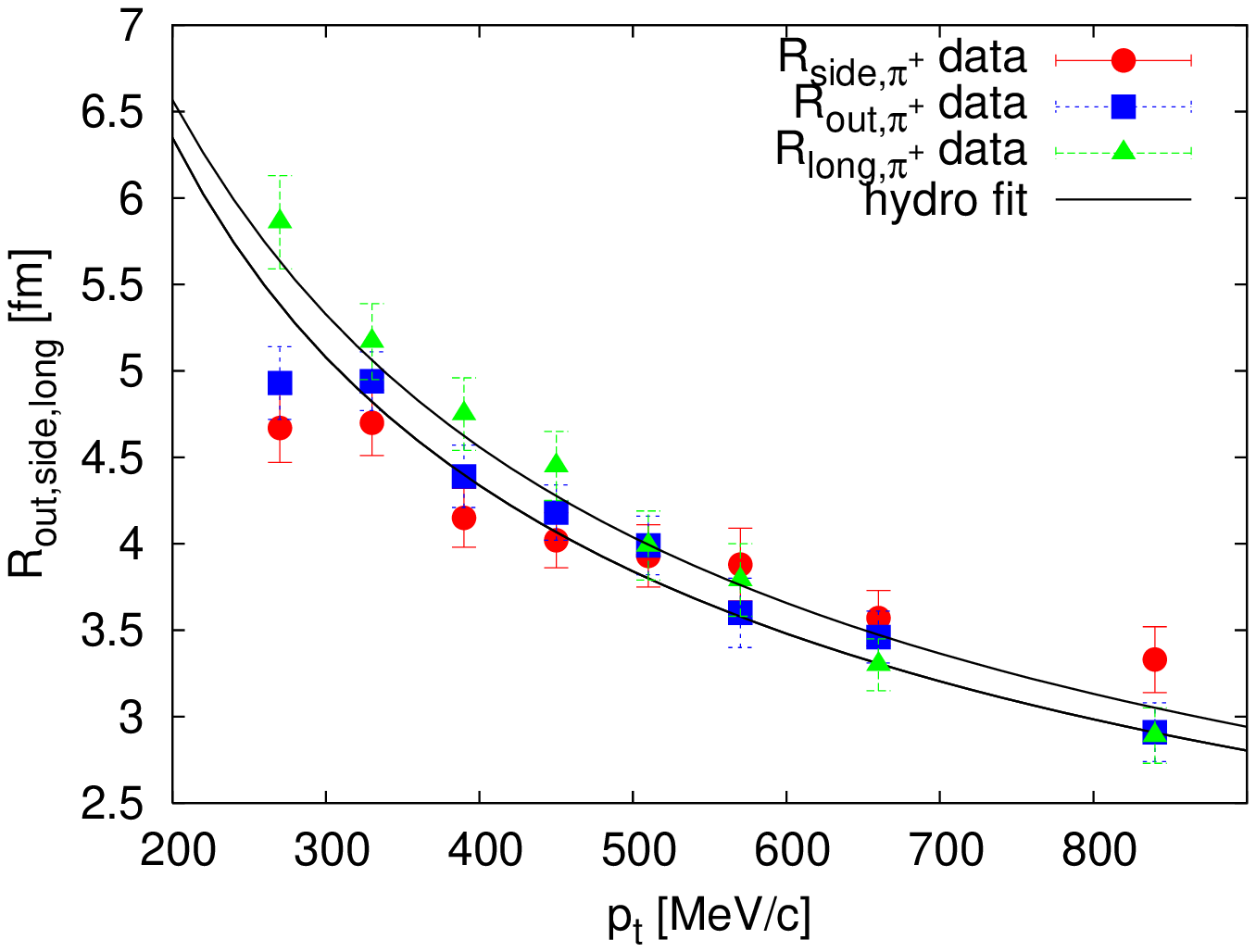}
		\includegraphics[height=0.48\textwidth,angle=270]{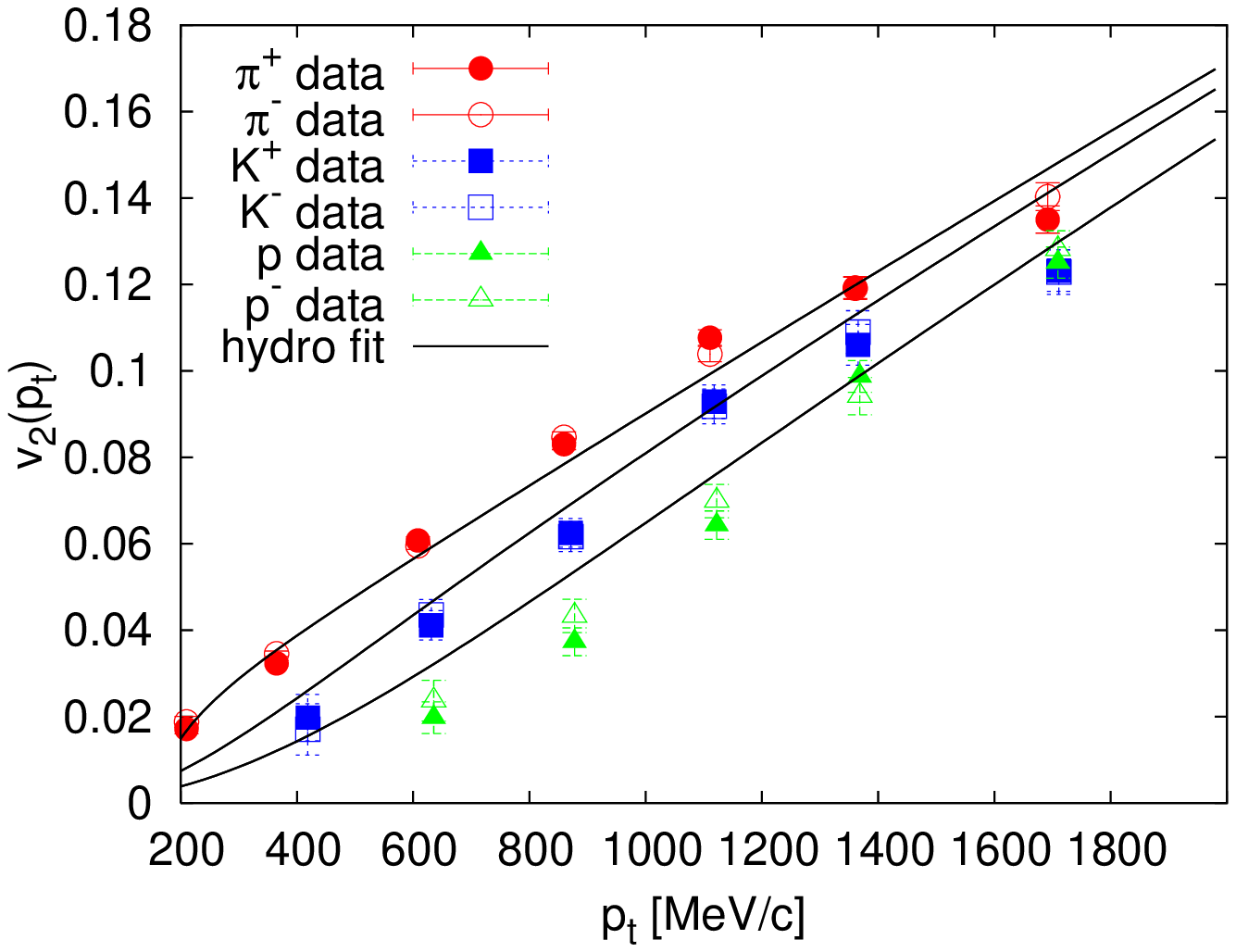}
	\caption{Fits to 0-30\% centrality PHENIX Au+Au spectra~\cite{Adler:2003cb} (top) HBT radii~\cite{Adler:2003kt} (middle) and 0-92\% centrality PHENIX Au+Au elliptic flow~\cite{Adler:2004rq} (bottom). See the obtained parameters in table~\ref{t:pars}. In the middle plot the lower curve is the fit to $R_{out}$ and $R_{side}$.}
	\label{f:fits}
\end{figure}

The fit results are shown in fig.~\ref{f:fits}. The fit parameters are listed in table~\ref{t:pars}. The central freeze-out temperature
$T_0$ is around 200 MeV for both datasets, and the fireball is colder away from the center. The expansion eccentricity $\epsilon$ being positive tells us that
the expansion is faster in-plane. Because of the Hubble-flow this means that the source is in-plane elongated, similarly to the result of ref.~\cite{Csanad:2008af}.
The freeze-out happens at a proper-time of $\tau_0 = 7.7$ fm$/c$.

\begin{figure}
	\centering
		\includegraphics[width=0.48\textwidth]{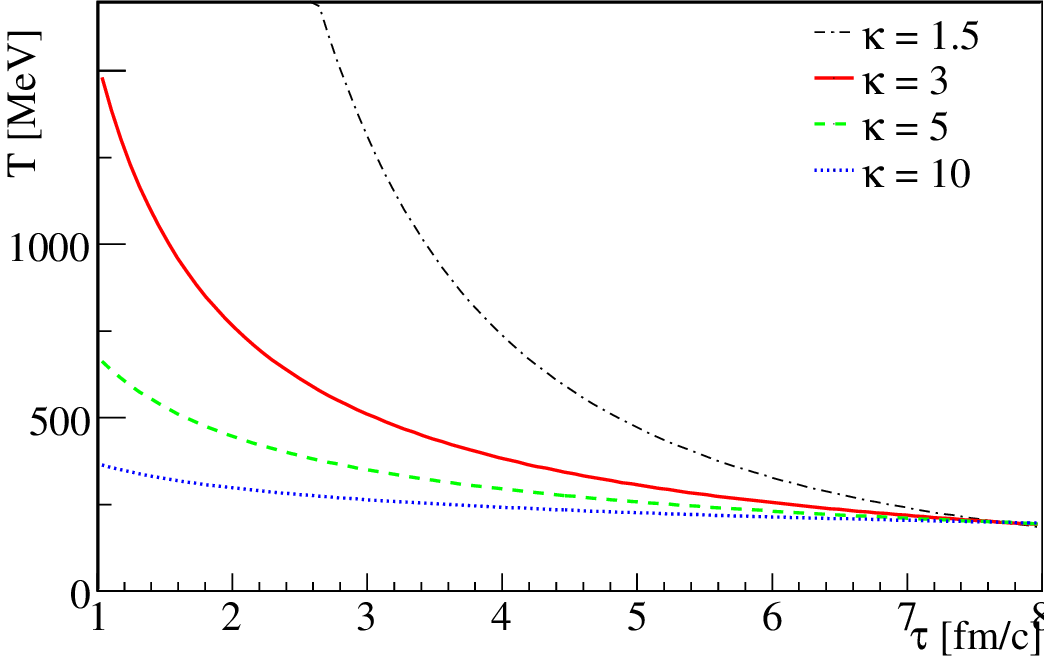}
		\includegraphics[width=0.48\textwidth]{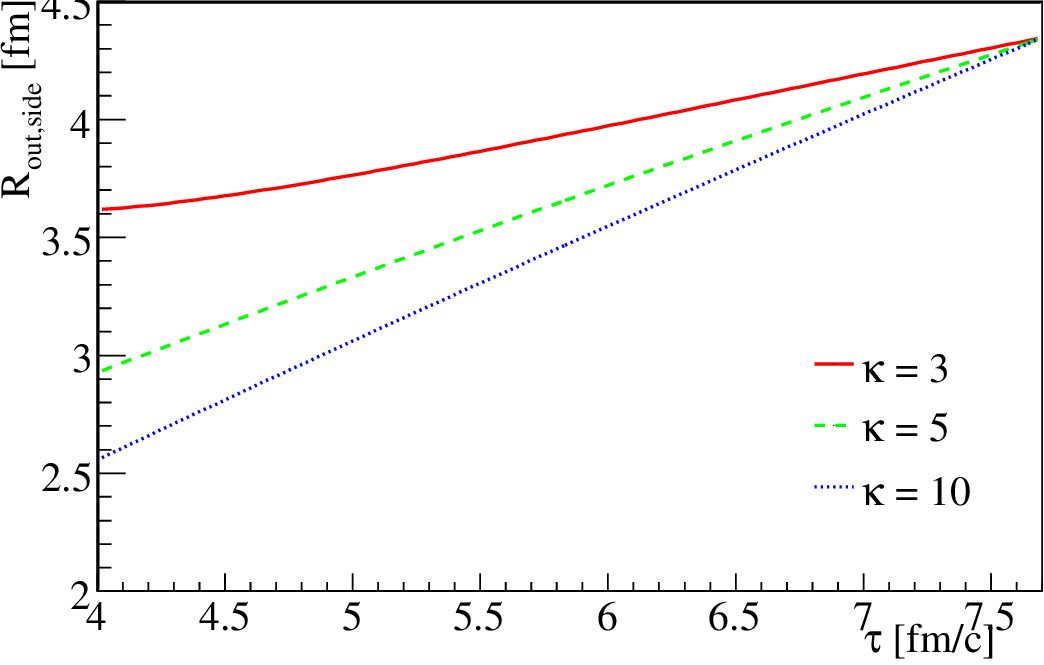}
		\includegraphics[width=0.48\textwidth]{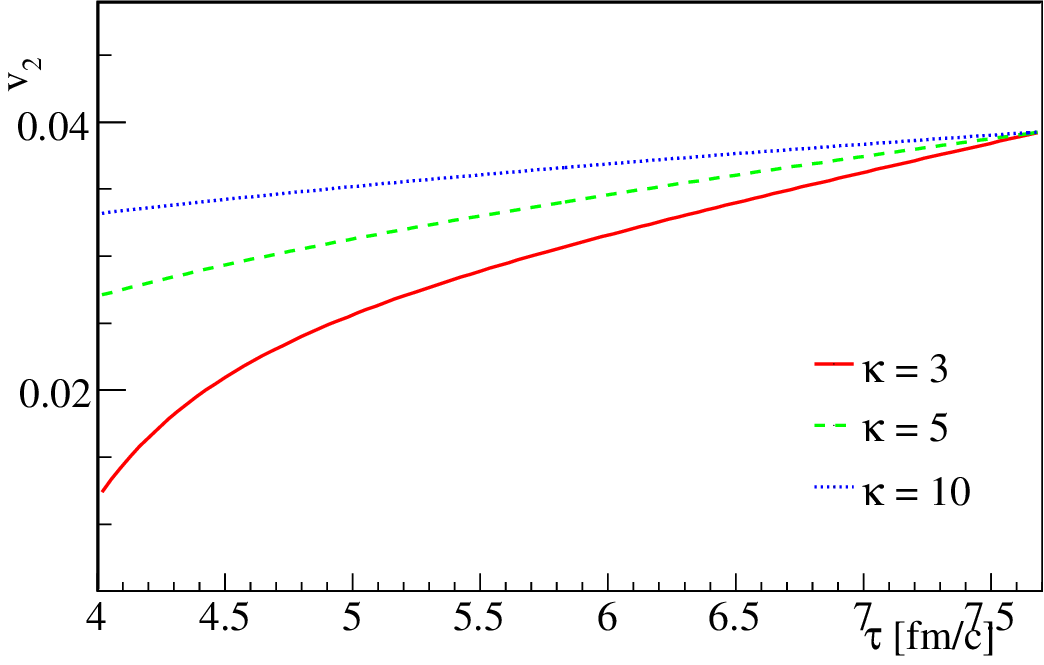}
	\caption{Time dependence of the central temperature of the fireball, from eq.~(\ref{e:temp}) (top) and time dependence of the HBT
	radii (middle) and the elliptic flow (bottom) where time dependence is included in the temperature $T_0$. The plots are shown for
	different $\kappa$ values. In reality $\kappa$ may change with time, we show here the curves only for fixed $\kappa$ values. Assuming an
	average of $\kappa=10$~\cite{Issah:2006qn} onegets an inital temperature of 370 MeV at $t_{\rm initial}$=1 fm$/c$, in agreement
	with PHENIX measurements~\cite{Adare:2008fqa}. The parameters used in this figure are that of table~\ref{t:pars}. Additionally a
	transverse momentum $p_t$ had to be specified to plot $v_2$, $R_{\rm out}$ and $R_{\rm side}$ at, we chose $p_t$ = 400 MeV/$c$.
	Note furthermore that as we go backwards in time, the central temperature reaches a point where the $E>T_0$ criterion is not valid
	any more. Thus in $v_2$, $R_{\rm out}$ and $R_{\rm side}$ we can go back only while $T<E$ is valid, i.e. roughly
	4 fm/$c$ for $\kappa=3$ at the given $p_t$. We did not plot the $\kappa=1.5$ case on those plots as the validity ends there at much bigger times. }
	\label{f:temp}
\end{figure}

Our fit parameters describe the fireball at the freeze-out. However, the solution is time-dependent, most importantly the temperature
depends on time as described by eq.~(\ref{e:temp}). We plotted the time-dependence of the central temperature in fig.~\ref{f:temp} for several
values of $\kappa$, i.e. several EoS'. From this, assuming for example an average $\kappa$ of 10~\cite{Issah:2006qn} one can also calculate
the initial central temperature of the fireball based on eq.~(\ref{e:temp}):
\begin{align}
T_{\rm initial} = T_0 \left(\frac{\tau_0}{\tau_{\rm initial}}\right)^{3/\kappa}
\end{align}
This yields 370 MeV at $t_{\rm initial}$=1 fm$/c$ (note that $t=\tau$ at the center of the fireball),
which is in agreement with PHENIX measurements~\cite{Adare:2008fqa}.

All the hadronic observables depend on the time of the freeze-out through the central temperature parameter $T_0$. If the freeze-out happens earlier or later,
the central temperature is larger, and this enters into the HBT radii and the elliptic flow as well through the centeral freeze-out temperature. However,
if we fix these final state (freeze-out) parameters and assume a $\kappa$ value, we can go back in time using the solution. We have to take eqs.~(\ref{e:v2})
and (\ref{e:ros}) and substitute the time dependent central temperature:
\begin{align}
T_0 \rightarrow T_0 \left(\frac{\tau_0}{\tau}\right)^{3/\kappa}.
\end{align}
See such plots in fig.~\ref{f:temp}, for different values of $\kappa$. In this figure we plot $v_2$ and $R_{\rm out}=R_{\rm side}$ for illustration purposes
at $p_t=400$ MeV/$c$, the other parametes are taken from the fits. Note that in reality $\kappa$ changes over time, i.e. it goes smoothly to the EoS of
a collisionless hadron gas.

\section{Summary}
Exact parametric solutions of perfect hydrodynamics were long searched for in order to describe the matter produced in heavy ion collisions at RHIC.
We extracted observables for the first time from the relativistic, 1+3 dimensional, ellipsoidally symmetric, exact solution of ref.~\cite{Csorgo:2003ry}.
We calculated momentum distribution, elliptic flow and Bose-Einstein correlation radii from the solution. We compared the results to 200 GeV Au+Au
PHENIX data~\cite{Adler:2003cb,Adler:2003kt,Adler:2004rq}. The solution is compatible with the data. The fitted parameters of the solution
describe the hadronic freeze-out. In the framework of our solution the fireball is in-plane elongated and has a Gaussian temperature profile. 
If using an experimentally determined average EoS of $\kappa\approx10$~\cite{Issah:2006qn}, our results yield approximately 370 MeV at
$\tau_{\rm initial}$=1 fm$/c$, in agreement with recent PHENIX measurements~\cite{Adare:2008fqa}.

\section*{Acknowledgments}
Supported by a bilateral collaboration between Hungary and Slovakia under project SK/20/2006 (Hungary) SK-MAD-02906 (Slovakia).
The authors gratefully acknowledge the support of the Hungarian OTKA grant NK 73143. They also would like to thank T.~Cs\"{o}rg\H{o} for
valuable discussions.

\bibliographystyle{prlstyl}
\bibliography{Master}

\end{document}